\documentclass[twocolumn]{emulateapj}
\usepackage{natbib}
\usepackage{graphicx}
\usepackage[dvipsnames]{color}
\usepackage{float}
\usepackage{longtable} 
\begin{document}
\title
{Proton Synchrotron Radiation from Extended Jets of PKS 0637-752 and 3C 273}
\author{Wrijupan Bhattacharyya\footnote{wriju.phys@gmail.com}, Nayantara Gupta}
\affil{Raman Research Institute, C. V. Raman Avenue, Sadashivanagar, 
Bangalore 560065, India}

\begin{abstract}
Many powerful radio quasars are associated with large scale jets, exhibiting bright knots as shown by high resolution images from the Hubble Space Telescope and Chandra X-ray Observatory. The radio\---optical flux component from these jets can be attributed to synchrotron radiation by accelerated relativistic electrons while the IC/CMB model, by far has been the most popular explanation for the observed X-ray emission from these jets.
Recently, the IC/CMB X-ray mechanism has been strongly disfavoured for 3C 273 and PKS 0637-752 since the anomalously hard and steady gamma-ray emission predicted by such models violates the observational results from Fermi-LAT.
Here we propose the proton synchrotron origin of the X ray\---gamma ray flux from the knots of PKS 0637-752 with a reasonable budget in luminosity, by considering synchrotron radiation from an accelerated proton population. Moreover, for the source 3C 273, the optical data points near 10$^{15}$ Hz could not be fitted using electron synchrotron (Meyer et al. 2015). We propose an updated proton synchrotron model, including the optical data from Hubble Space Telescope, to explain the common origin of optical\---X-ray\---gamma-ray emission from the knots of quasar 3C 273 as an extension of the work done by Kundu $\&$ Gupta (2014).
We also show that TeV emission from large scale quasar jets in principle, can arise from proton synchrotron, which we discuss in the context of knot wk8.9 of PKS 0637-752.
\end{abstract}

\keywords{quasars:general; X-rays: galaxies; gamma-rays:galaxies}

\section{Introduction}

Quasars are a class of active galaxies, which are highly energetic and powered by accretion of mass around central supermassive
black-holes. Jets of FRII radio galaxies and quasars often exhibit regions of extreme brightness or knots, as have been 
observed by radio telescopes
since a long time back. But more recently, images
from \textit{Hubble Space Telescope (HST)} and \textit{Chandra X-Ray Observatory} have shown that significant amount of high energy
radiation is also produced from the bright knots present in these jets. The first discovery of high-energy X-ray emission from 
the kpc-scale relativistic jet of quasar PKS 0637-752 by the Chandra Observatory (Chartas et al. (2000)) has led to many similar 
significant discoveries later (Harris \& Krawczynski (2006)).
\par

PKS 0637-752, located at redshift z = 0.651 (Savage,Browne and Bolton (1976)) was the first X-ray target of the Chandra Observatory 
(Weisskopf et al. (2000); Schwartz et al. (2000); Chartas et al. (2000)) which accidentally discovered a 100 kpc one-sided jet from 
the source coincident with the radio jet reported by Tingay et al. (1998). The strong X-ray flux observed from this jet is
difficult to explain via standard mechanisms such as Synchrotron Self Compton (SSC) or thermal Bremsstrahlung (Schwartz et al.
2000). Since the launch of \textit{Chandra} in 1999, many tens of such quasar jets luminous in X-rays have been detected. The periodic structure in knots in the megaparsec scale jet of PKS0637-752 has been observed by the Australian 
Telescope (Godfrey et al. (2012)).
\par
3C 273, located at redshift z=0.158 (Strauss et al.(1992)) is the brightest and 
 most studied AGN and is accompanied by a large scale jet of projected length 57 kpc (Harris $\&$ Krawczynski (2006)). 
Analysis has been performed on its broadband energy spectrum by various authors (Jester et al. (2001), (2005); Uchiyama et al. (2006); Meyer $\&$ Georganopulos (2014)). Soldi et al. (2008) have analyzed long-term multiwavelength data to study its temporal variability properties.
Sambruna et al. (2001) and Uchiyama et al. (2006) have found two distinct components in the jet emission of 3C 273. 
The radio emission from the kpc scale jet has been explained with synchrotron radiation from shock-accelerated relativistic electrons by Marscher $\&$ Gear (1985) and T\"urler et al. (2000). 
  
 \par

The radio to optical spectral energy distribution from jets of both 3C 273 and PKS 0637-752, can be well explained by synchrotron radiation of
relativistic electrons present in the jets (Samburna et al. (2001)). However in several jets, the X-ray emission
from the knots is much higher and/or harder than expected from the radio-optical synchrotron spectrum as explained by Schwartz et al. (2000) for quasar PKS 0637-752. Synchrotron self-Compton (SSC) mechanism could not explain the observed 
 X-radiation in both the sources, as explained by Chartas et al. (2000) for PKS 0637-752 and Samburna et al. (2001) for 3C 273.
Hence it was suggested by Tavecchio et al. (2000) and Celotti et al. (2001), 
that the X-ray emission from PKS 0637-752 can arise due to inverse Compton scattering of the cosmic microwave background photons by shock-accelerated relativistic electrons in the jet (IC/CMB). The IC/CMB model has also been applied to explain X-ray radiation from Knot A of 3C 273 (Samburna et al. (2004)).
IC/CMB mechanism by far has been the most popular explanation for X-radiation in quasar jets. Samburna et al. (2004) explained
the X-ray emission from most of the radio jets included in their survey, by IC/CMB. IC/CMB model predictions was verified for 
the source PKS 1150+49 by follow-up observations by Samburna et al. (2006). \par

However, a number of problems have been noticed with the IC/CMB X-ray model.
It requires the jet to remain highly relativistic (i.e. high Lorentz factor$\sim$10-20) even upto kpc scales and pointed at a very small angle to our line 
of sight. But it was predicted by Arshkian $\&$ Longair (2004) that the jets decelerate and can only remain mildly relativistic when they reach kpc scales, although this lacks direct experimental support. 
Moreover, the small angle subtended by the jet to our line of sight sometimes gives rise to deprojected jet length of Mpc size. Also IC/CMB models require huge jet kinetic power, sometimes exceeding the Eddington limit for the source (Uchiyama et al. (2006)). As an alternative to IC/CMB, it was proposed that X-radiation
from large scale quasar jets can also be explained by synchrotron radiation from a second shock accelerated electron population, different from the one giving rise to the radio-optical spectra (Jester et al. (2006), Uchiyama et al. (2006)). 
Although this second synchrotron model overcomes the problem of super-Eddington power requirements, high Lotentz factor and Mpc scale jet length of the IC/CMB to explain the observed X-Ray data, the problem lies in its unexplained co-spatial existence with the first high energy electron population (Schwartz et al. (2000)). \par

Recently it has been shown by Meyer et al. (2014 and 2015) using observational results from Fermi-LAT, that IC/CMB incorrectly predicts 
the gamma-ray flux at GeV energies. They have shown using long term Fermi monitoring data that the hard and steady gamma-ray
emission implied by the IC/CMB X-ray models, overproduces the GeV flux thus violating observational results from Fermi 
for both quasar jets PKS 0637-752 and 3C 273. Thus IC/CMB is ruled out as possible X-ray emission mechanism in both of our target
sources. 
The implication of explaining X-ray emission from the knots with electron synchrotron also has been discussed by Meyer et al. (2015). The shock accelerated electrons emitting X-rays in synchrotron emission would also give GeV-TeV gamma rays by inverse Compton scattering off the CMB photons. The luminosity expected in TeV gamma rays is very high in this case.

\par
Aharonian (2002) proposed synchrotron radiation by a shock-accelerated proton population which explained the radio to X-ray spectrum from knot A of 3C 273.  Also the jet emission from PKS 0637-752 in optical to X ray spectrum was well explained by the proton synchrotron model in this paper. A broken power law spectrum of accelerated protons having energy upto $10^{20}$ eV was used, where the spectral indices were determined by three important time scales the synchrotron loss and escape time scales of the protons and the age of the jet. The protons lose energy very slowly in the magnetic field of order milli Gauss in the jet as a result they can diffuse through the length of the kpc scale jet. 
\par

We have used this proton synchrotron model (Aharonian 2002) to propose the possible common origin of the high energy photons in kpc scale jets of quasars PKS 0637-752 and 3C 273. For the source PKS 0637-752 we have modeled the X-ray\---gamma-ray flux by proton synchrotron. Kundu $\&$ Gupta (2014) demonstrated the possible proton synchrotron origin of X-ray and gamma-ray emission from the large scale jet of 3C 273. In the recent work of Meyer et al. (2015), the optical HST data near 10$^{15}$ Hz for the source 3C 273 has not been included in the radio-optical synchrotron fit, which we have included in our updated proton 
synchrotron model proposing a common origin of optical, X-ray and gamma-ray photons. \par 
We have also discussed about the possibility of TeV photon emission from large scale quasar jets within the proton synchrotron model. 
For the knot wk8.9 of PKS 0637-752, we have shown that proton synchrotron mechanism can in principle, give rise to a TeV flux within a reasonable budget in luminosity if protons are accelerated to energy close to 10$^{21}$ eV. 
\section{The Proton Synchrotron Model}
\label{sect:model}

\subsection{High Energy Spectral Energy Distribution}
We describe the formalism used in our work (earlier discussed in Aharonian 2002, Kundu \& Gupta 2014). The shock accelerated protons are diffusing through the large scale jets of the quasars and losing energy due to synchrotron emission and diffusion. We have calculated the Doppler factors ($\delta_D$) of the jets assuming their Lorentz factor to be $\Gamma=3$ to fit the observational data.
Within a spherical blob of size $R$ and magnetic field $B$ the relativistic protons
 are trapped. Their escape time scale is 
\begin{equation}
t_{\rm esc}\simeq 4.2\times 10^5 \eta^{-1} B_{\rm mG} R^{2}_{\rm kpc} (E/10^{19}\rm eV)^{-1} {\rm yr}.
\label{bohm_esc}
\end{equation}  
In Bohm diffusion limit the gyrofactor $\eta=1$.
Another expression for the escape time which is energy dependent reduces the energy budget (Aharonian 2002)

\begin{equation}
t_{\rm{esc}}=\frac{1.4\times 10^7}{(E/10^{14}\rm {eV})^{0.5}} \rm {yr}.
\label{time_esc}
\end{equation}

The synchrotron energy loss time scale of the relativistic protons in the 
jet is
\begin{equation}
t_{\rm{synch}}\simeq 1.4\times 10^{7} B^{-2}_{\rm{mG}} (E/10^{19} \rm eV)^{-1} \rm{yr}.
\label{synch_time}
\end{equation} 

We have considered the broken power law spectrum of the shock accelerated 
 relativistic protons
\begin{equation}
\frac{dN_{p}(E_p)}{dE_p} 
=A \left\{ \begin{array}{l@{\quad \quad}l}
{E_p}^{-p_1} &
E_p<E_{p,br}\\{E_{p,br}}^{(p_2-p_1)}
{E_p}^{-p_2} & E_p>E_{p,br}.
\end{array}\right.
\end{equation}
We compare the synchrotron loss and escape time scales of the protons in the jet whose age is assumed to be 3$\times 10^{8}$ years. When the synchrotron loss time scale is shorter than the escape or diffusion time scale given in eqn (\ref{bohm_esc}) and the age of the jet, synchrotron loss becomes important. As a result the spectrum of high energy protons steepens by $E_p^{-1}$ above the break energy $E_{p,br}$ (our model 1).
For knot wk8.9 of PKS 0637-752 at $E_{p,br}$=10$^{16}$ eV, the synchrotron loss time scale is $t_{sync}$ = 1.4$\times10^{8}$
yrs, which is smaller than the age of the jet and the escape time ($t_{esc}$=7.56$\times10^{9}$ yrs); similar is the case
in the combined knot scenerio, thus increasing the spectral index by 1. For knot A and knots A+B1 combined of 3C 273, at $E_{p,br}$=10$^{16}$ eV (and 5.62$\times10^{15}$ eV), 
$t_{sync}$ = 1.4$\times10^{8}$ yrs (and 2.49$\times10^{8}$ yrs) whereas $t_{esc}$ = 1.52$\times10^{10}$ yrs and 8.14$\times10^{10}$ yrs
respectively. \par
We have also considered another scenario where the escape time scale given in eqn(\ref{time_esc}) becomes shorter than the synchrotron time scale and the age of the jet for very high energy protons. This results in a steeper spectrum by a factor of $E_p^{-0.5}$ above the break energy $E_{p,br}$ (our model 2).
For PKS 0637-752 both for the single knot and combined knot scenarios, at $E_{p,br}$=10$^{12}$ eV, the escape time is
1.4$\times10^{8}$ yrs which thus becomes dominant over the synchrotron loss ($t_{sync}\sim10^{12}$ yrs) and age of the jet. 
Similar is the case for knot A and the combined knots of 3C 273, where also according to our model 2, the escape loss becomes
more important compared to other time scales.

\par
The high energy photon spectrum from knots of PKS 063-752 and 3C 273 are compared with the theoretical predictions 
of our model 1 and model 2 (see Fig.1. to Fig.4.). The values of the parameters used in our flux calculations are given 
in Table 1 and 2.

\subsection{Proton Synchrotron Origin of TeV Gamma-rays}
In this section we show that proton synchrotron radiation can in principle, give rise to TeV gamma-rays from extended quasar jets within a reasonable budget in photon luminosity. TeV blazars can exhibit luminosity beyond $10^{42}$ erg/sec (Abramowski et al. (2014)). 
We discuss the implications of our proton synchrotron model in the context of TeV emission, for knot wk8.9 of PKS 0637-752. The expressions for escape and synchrotron time scales used in our models are given in eqn (\ref{time_esc}) and eqn (\ref{synch_time}). For the range of parameters considered, the synchrotron loss becomes shorter than escape loss and jet age resulting in a steeper proton spectrum by $E_{p}^{-1}$ above the break energy. We propose that proton synchrotron radiation can give rise to TeV emission from knot wk8.9, if proton acceleration to energies near $10^{21}$ eV is possible. 
Ebisuzaki $\&$ Tajima (2013) have discussed that protons/nuclei in AGN jets can be accelerated to beyond $10^{21}$ eV by plasma wakefield field formed by intense electromagnetic field.
Our parameters estimates are listed in Table 3 and the model fits can be found in Fig.5. The models have been constructed under the assumption of equipartition in energy density of the particles and magnetic field. 
In our calculations we have shown the intrinsic source spectrum, not the observed spectrum. However, due to severe absorption of the TeV gamma-rays by the extragalactic background light (EBL), direct observation of the TeV spectra would be difficult. Our photon luminosity budget near 1 TeV, for the four models considered $\sim10^{41}-10^{43}$ erg/sec, which is reasonable for TeV blazars, thus implying the validity of our proposition.

\section{Results and Conclusion}
The radio to optical data from the single knot wk8.9 and the knots wk7.8, wk8.9,wk9.7,wk10.6 of PKS 0637-752, the single knot A and the combined knots A, B1 of 3C 273 are fitted by the synchrotron emission of shock accelerated electrons by Meyer et al. 2015. We have fitted the higher energy photon data (X-ray data and Fermi LAT upper limits) from these knots with proton synchrotron mechanism. Moreover,
 in the case of 3C 273 the optical data at $10^{15}$ Hz which cannot be fitted by electron synchrotron emission (Meyer et al. 2015) has been included within our updated proton synchrotron models.
The existence of very high energy protons ($\sim 10^{20}$ eV) in the kpc-scale knots, is the basic assumption of this model. \par

In the work of Aharonian (2002), it was proposed that proton-synchrotron can give rise to radio to X-ray flux from extended quasar jets. In this work it was assumed that during the jet lifetime, protons with a time independent energy spectrum are injected (quasi) continuously into a spherically symmetric blob. Aharonian considered three models to explain the observed spectral energy distribution, each of which reduces the energy budget compared to the previous one.
For Knot A of 3C 273 with jet lifetime 3$\times10^7$ yrs, the first model uses a broken power-law spectrum of protons with spectral indices 2.4 and 3.4 below and above the break and a magnetic field B = 5 mG. This model fits the radio to X-ray data with luminosity in magnetic field $L_B$= 1.33$\times10^{44}$ erg/s and that in protons $L_p$= 1.2$\times10^{47}$ erg/s, implying a large deviation from equipartition. To reduce the energy budget Aharonian's second model considers magnetic field B = 10mG, initial proton spectral index p1 = 2, an energy-dependent escape time scale which becomes dominant over synchrotron losses, thus resulting in spectral index p2 = 2.5 after break. This model also explains the radio to X-ray spectrum of 3C 273, but for reduced energy requirements ($L_B=1.1\times10^{45}$ erg/s; $L_p=1.1\times10^{44}$ erg/s).To further reduce the luminosity, in his third model Aharonian adopted a power-law spectrum with exponential cut-off at E = 10$^{18}$ eV, which fits only the X-ray data for a magnetic field value of 3 mG and spectral index 2. The luminosities in cosmic ray protons and magnetic field are $L_p$=10$^{45}$ erg/s and $L_B$=3.7$\times10^{44}$ erg/s respectively, which is less compared to the other two models.\par 

For PKS 0637-752 instead of taking the individual knots, Aharonian considered that the overall X-ray emission is coming from a single source. The first model considers a broken power law with exponential cut-off at E = 10$^{20}$ eV and spectral indices 1.75 and 2.75 below and above break. It was assumed that particles propagate in relaxed-Bohm diffusion limit with magnetic field 1.5 mG and size of emitting region 5 kpc, where escape losses dominate over synchrotron losses. This model fits the optical to X-ray spectrum of PKS 0637-752 with a large proton acceleration power $L_p$=3$\times10^{46}$ erg/s. The second model adopted a higher value of magnetic field (B=3 mG), size was reduced to 3kpc and it was assumed that particle propagation takes place in the Bohm regime which results in dominance of the synchrotron loss time scale over escape time scale. In order to reduce the proton acceleration power by another order of magnitude, in his third model Aharonian uses an early exponential cut-off at E=2$\times10^{18}$ eV, which requires a proton power of 2.9$\times10^{45}$ erg/s. \par 

In our work, we revisit the proton synchrotron models to explain the higher energy observations from knots of PKS 0637-752 and 3C 273, under the assumption that particles diffuse in the Bohm limit ($\eta$=1). We fit the X-ray to gamma-ray observational data from Knot wk8.9 and combined knots wk7.8, wk8.9, wk9.7, wk10.6 of PKS 0637-752 with parameter estimates according to Table 1.
The spectral index of protons has been varied in the range of 1.35-1.9 below the break in model 1 and 2 and luminosities required to explain the observed X-ray to gamma ray spectral energy distribution from the knots of PKS 0637-752 are $\sim 6\times10^{43}$ erg/s which is about 0.6$\%$ of Eddington luminosity of the source (see Table 1). 
Our models also explain the optical to gamma-ray energy spectrum from Knots A and A+B1 of 3C 273 with luminosity $\sim5\times10^{43}$ erg/s in model 1 and 2 (0.5$\%$ of Eddington luminosity)(see Table 2). For this source in order to match the experimental observations, we consider an initial proton spectra in the range p1 = 1.57-2.03 which changes by 1 or 0.5 (according to model 1 or model 2) after break. In all cases our model 2 fits the observed photon data with equipartition of energy between the magnetic field and the relativistic cosmic-ray protons. Thus our model 2 remains more favorable.\par

Also we discuss the possible proton synchrotron origin of TeV component from extended quasar jets, in the context of knot wk8.9 of PKS 0637-752 and show that TeV emission is in principle possible with photon luminosity $10^{41}-10^{43}$ erg/s at the peak near 1 TeV in the energy spectrum, if protons in the kpc-scale jets are accelerated upto 5.6$\times10^{21}$ eV. However direct observation of such TeV spectrum would be difficult due to severe EBL absorption.

\begin{table*}[ht]
\centering
\tablecaption{}{TABLE 1 : Model parameters for PKS 0637-752}\\
\scalebox{1.4}{
\begin{tabular}{|c|c|c|c|c|}
\hline
Knot  & Parameter & Notation & Model 1  & Model 2\\
\hline
wk8.9    &Size of knot (m) &R &2.2$\times$10$^{19}$ &3.6$\times$10$^{19}$\\
         &Lorentz factor &$\Gamma$ &3 &3\\
         &Viewing angle &$\theta$ & 35$^\circ$ &30$^\circ$\\
         &Doppler factor &$\delta_D$ &1.46 &1.79\\
         &Magnetic field (mG) &B &8 &5\\
         &Minimum proton energy(eV) &E$_{p,min}$ &10$^{14}$ &10$^{10}$\\
         &Maximum proton energy(eV) &E$_{p,max}$ &7.2$\times10^{19}$ &5.2$\times10^{19}$\\
         &Break proton energy(eV) &E$_{p,br}$ &10$^{16}$ &10$^{12}$\\
         &Low energy proton spectral index &p1 &1.35 &1.63\\
         &High energy proton spectral index &p2 &2.35 &2.13\\
         &Luminosity in magnetic field (erg/sec) &L$_{B}$ &1.19$\times$10$^{43}$ &2.06$\times$10$^{43}$\\
         &Luminosity in proton (erg/sec) &L$_{p}$ &7.07$\times$10$^{42}$ &2.06$\times$10$^{43}$\\
\hline
wk7.8+    &Size of knot (m) &R &2.1$\times$10$^{19}$ &4.6$\times$10$^{19}$\\
wk8.9+         &Lorentz factor &$\Gamma$ &3 &3\\
wk9.7+         &Viewing angle &$\theta$ & 35$^\circ$ &23$^\circ$\\
wk10.6         &Doppler factor &$\delta_D$ &1.46 &2.47\\
         &Magnetic field (mG) &B &9 &7\\
         &Minimum proton energy(eV) &E$_{p,min}$ &10$^{14}$ &10$^{10}$\\
         &Maximum proton energy(eV) &E$_{p,max}$ &5.62$\times10^{19}$ &5.2$\times10^{19}$\\
         &Break proton energy(eV) &E$_{p,br}$ &1.58$\times$10$^{16}$ &10$^{12}$\\
         &Low energy proton spectral index &p1 &1.35 &1.9\\
         &High energy proton spectral index &p2 &2.35 &2.4\\
         &Luminosity in magnetic field (erg/sec) &L$_{B}$ &1.3$\times$10$^{43}$ &8.4$\times$10$^{43}$\\
         &Luminosity in protons (erg/sec) &L$_{p}$ &1.3$\times$10$^{43}$ &8.4$\times$10$^{43}$\\
\hline
\end{tabular}}
\vspace{0.2 cm}
\end{table*}

\begin{figure*}[htp]
\centering
\includegraphics[scale=1.]{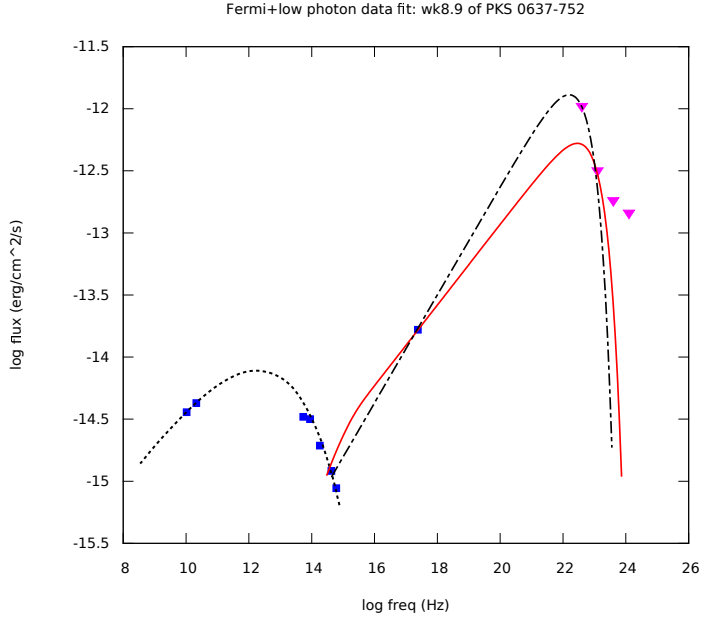}
\caption{The SED of knot wk8.9 of PKS 0637-752. The radio to X-ray data (Mehta et al. (2009),Meyer et al. (2015))
are shown by blue points and the Fermi-LAT upper 
limit (Meyer et al. (2015)) shown by magenta arrows. Black dotted line: Electron synchrotron fit to radio-optical data (Meyer et al. (2015)); 
Red solid line: Proton synchrotron fit according to our Model 1 as in Table 1; Black dot-dashed line: Our Model 2. The code used
in our work is from Krawczynski et al. (2004)}
\label{WK89}
\end{figure*}
\begin{figure*}[htp]
\centering
\includegraphics[scale=1.]{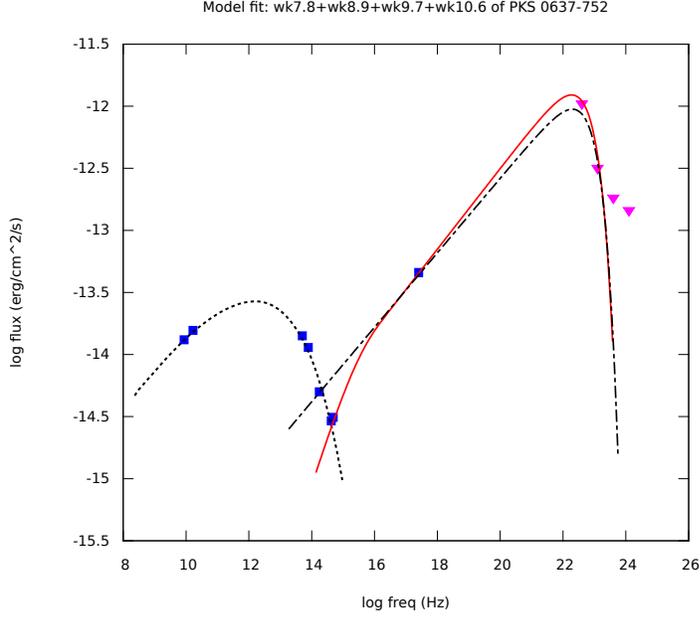}
\caption{The SED of combined knots wk7.8+wk8.9+wk9.7+wk10.6 of PKS 0637-752. Fermi-LAT upper limits, radio to X-ray
data and model fits follow the same notation as Fig. 1 (references as described in figure 1)}
\label{comb}
\end{figure*}

\begin{table*}[ht]
\centering
\tablecaption{}{TABLE 2 : Model parameters for 3C 273}\\
\scalebox{1.4}{
\begin{tabular}{|c|c|c|c|c|}
\hline
Knot  & Parameter & Notation & Model 1  & Model 2\\
\hline
A    &Size of knot (m) &R &1.9$\times$10$^{19}$ &3.15$\times$10$^{19}$\\
         &Lorentz factor &$\Gamma$ &3 &3\\
         &Viewing angle &$\theta$ & $45^\circ$ &23$^\circ$\\
         &Doppler factor &$\delta_D$ &1 &2.47\\
         &Magnetic field (mG) &B &10 &9\\
         &Minimum proton energy(eV) &E$_{p,min}$ &10$^{14}$ &10$^{10}$\\
         &Maximum proton energy(eV) &E$_{p,max}$ &1.9$\times10^{20}$ &8.9$\times10^{19}$\\
         &Break proton energy(eV) &E$_{p,br}$ &10$^{16}$ &10$^{12}$\\
         &Low energy proton spectral index &p1 &1.62 &2.02\\
         &High energy proton spectral index &p2 &2.62 &2.52\\
         &Luminosity in magnetic field (erg/sec) &L$_{B}$ &1.2$\times$10$^{43}$ &4.45$\times$10$^{43}$\\
         &Luminosity in protons (erg/sec) &L$_{p}$ &7.59$\times$10$^{42}$ &4.45$\times$10$^{43}$\\
\hline
A+    &Size of knot (m) &R &3.3$\times$10$^{19}$ &3.6$\times$10$^{19}$\\
B1         &Lorentz factor &$\Gamma$ &3 &3\\
        &Viewing angle &$\theta$ & $45^\circ$ &22$^\circ$\\
        &Doppler factor &$\delta_D$ &1 &2.59\\
         &Magnetic field (mG) &B &10 &9\\
         &Minimum proton energy(eV) &E$_{min}$ &10$^{14}$ &10$^{10}$\\
         &Maximum proton energy(eV) &E$_{max}$ &1.25$\times10^{20}$ &7.24$\times10^{19}$\\
         &Break proton energy(eV) &E$_b$ &5.62$\times$10$^{15}$ &10$^{12}$\\
         &Low energy proton spectral index &p1 &1.57 &2.03\\
         &High energy proton spectral index &p2 &2.57 &2.53\\
         &Luminosity in magnetic field (erg/sec) &L$_{B}$ &6.3$\times$10$^{43}$ &6.65$\times$10$^{43}$\\
         &Luminosity in protons (erg/sec) &L$_{p}$ &1.38$\times$10$^{43}$ &6.65$\times$10$^{43}$\\
\hline
\end{tabular}}
\vspace{0.2 cm}
\end{table*}

\begin{figure*}[htp]
\centering
\includegraphics[scale=1.]{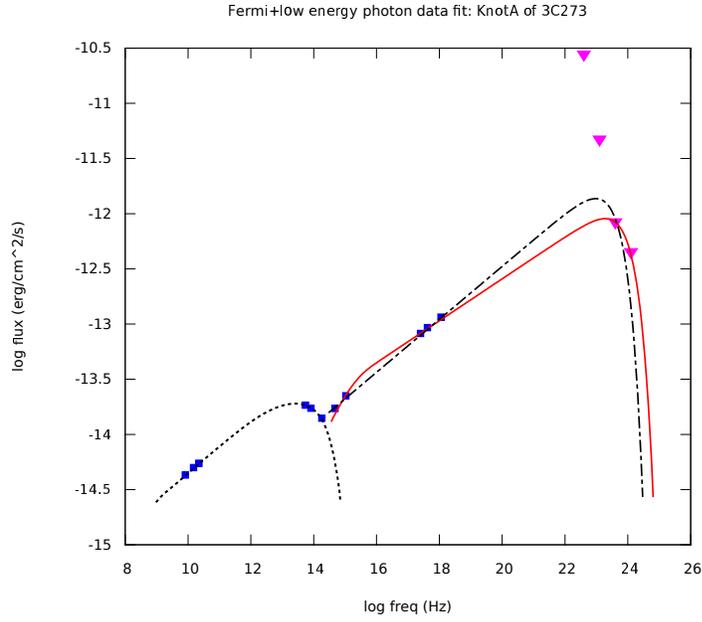}
\caption{The SED of knot A of 3C 273. Fermi-LAT upper limits (Meyer et al. (2015)) and radio to X-ray
data (Jester et al. (2005), Uchiyama et al. (2006)) are shown with magenta arrows and blue points. Black dotted line: Electron synchrotron 
spectrum (Meyer et al. 2015); Solid red line:
Our Model 1 as in Table 2; Black dot-dashed line: Our Model 2}
\label{knotA}
\end{figure*}

\begin{figure*}[htp]
\centering
\includegraphics[scale=1.]{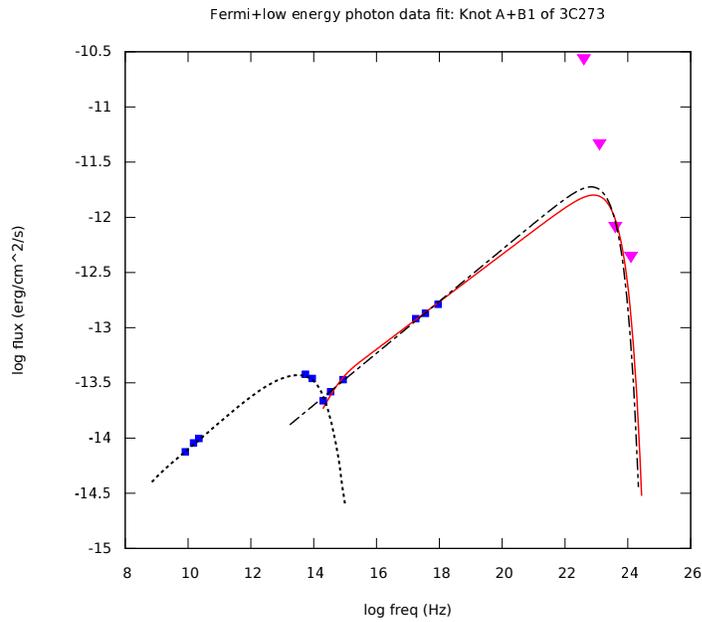}
\caption{The SED of combined knot A+B1 of 3C 273. Data references as in Fig. 3. Symbols and model fits follow same notation as in Fig.3.}
\label{knotAB1}
\end{figure*}
\begin{table*}[ht]
\centering
\tablecaption{}{TABLE 3: TeV emission from Knot wk8.9 of PKS 0637-752 at a distance of 3.8Gpc}\\
\scalebox{1.4}{
\begin{tabular}{|c|c|c|c|c|c|}
\hline
 Parameter &Notation & Model 1  & Model 2 & Model 3 & Model 4\\
\hline
Size of knot (m) &R &1.53$\times$10$^{19}$ &1.53$\times$10$^{19}$ &1.75$\times$10$^{19}$ &3.2$\times$10$^{19}$\\
         Lorentz factor &$\Gamma$ &3 &3 &3 &3\\
         Viewing angle &$\theta$ &28$^\circ$ &23$^\circ$ &23$^\circ$ &23$^\circ$\\
         Doppler factor &$\delta_D$ &1.9 &2.47 &2.47 &2.47\\
         Magnetic field (mG) &B &10 &8 &7 &5\\
         Min proton energy(eV) &E$_{min}$ &10$^{14}$ &10$^{14}$ &10$^{14}$ &10$^{14}$\\
         Max proton energy(eV) &E$_{max}$ &3.16$\times10^{21}$ &3.16$\times10^{21}$ &5.6$\times10^{21}$ &5.6$\times10^{21}$\\
         Break proton energy(eV) &E$_b$ &7.08$\times10^{15}$ &10$^{16}$ &1.99$\times10^{16}$ &3.98$\times10^{16}$\\
         Low energy proton spec. index &p1 &1.8 &1.9 &2 &2.25\\
         High energy proton spec. index &p2 &2.8 &2.9 &3 &3.25\\
   Luminosity in magnetic field (erg/sec) &L$_{B}$ &6.29$\times$10$^{42}$ &4.06$\times$10$^{42}$ &4.63$\times$10$^{42}$ &$1.44\times$10$^{43}$\\
   Luminosity in protons (erg/sec)        &L$_{p}$ &6.29$\times$10$^{42}$ &4.06$\times$10$^{42}$ &4.63$\times$10$^{42}$ &1.44$\times$10$^{43}$\\      
\hline
Photon luminosity in jet (erg/sec)        &L$_{j,ph}$ &1.81$\times$10$^{43}$ &6.74$\times$10$^{42}$ &2.75$\times$10$^{42}$ &2.28$\times$10$^{41}$\\
\hline
\end{tabular}}
\vspace{0.2 cm}
\label{}
\end{table*}
\begin{figure*}[htp]
\centering
\includegraphics[scale=1,angle=0]{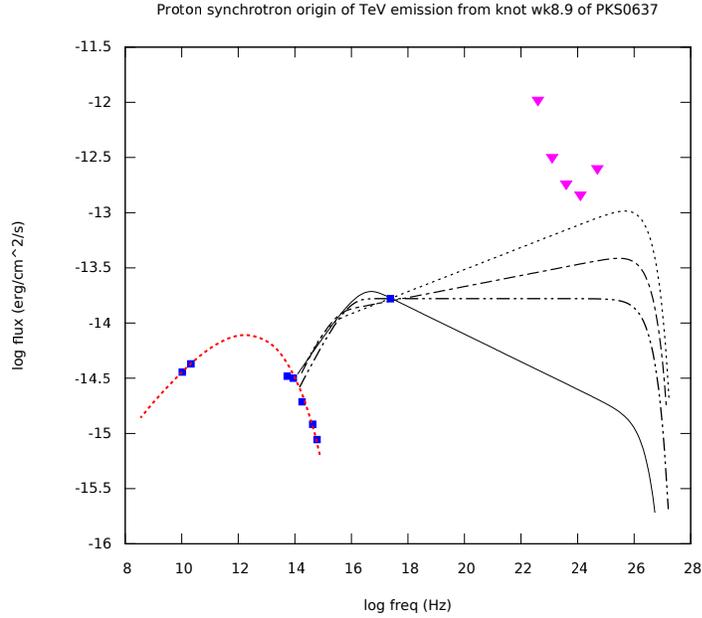}
\caption{Proton-synchroton modelling of TeV emission from Knot wk8.9 of PKS 0637-752. Fermi-LAT and lower energy photon data references and notation as in Fig. 1;.
Red dotted line: electron synchrotron spectrum; Black dotted line: Model 1 according to Table 3;
Dot-dashed line: Model 2;  Dot-dot-dashed line: Model 3; Solid line: Model 4}
\label{TeV}
\end{figure*}

\end{document}